\documentclass[%
 reprint,
superscriptaddress,
 amsmath,amssymb,amsfonts,
prl,
]{revtex4-2}

\usepackage{graphicx}
\usepackage{dcolumn}
\usepackage{bm, bbold}
\usepackage{comment, color}
\usepackage{natbib}
\usepackage{tabularx}

\begin{document}


\def\bea{\begin{eqnarray}}
\def\eea{\end{eqnarray}}
\def\beq{\begin{equation}}
\def\eeq{\end{equation}}
\def\f{\frac}
\def\k{\kappa}
\def\e{\epsilon}
\def\ve{\varepsilon}
\def\be{\beta}
\def\D{\Delta}
\def\h{\theta}
\def\t{\tau}
\def\a{\alpha}

\def\cDa{{\cal D}[X]}
\def\cD{{\cal D}[x]}
\def\cL{{\cal L}}
\def\cLo{{\cal L}_0}
\def\cLa{{\cal L}_1}
\def\rv{{\bf r}}
\def\tv{\hat t}
\def\on{{\omega_{\rm a}}}
\def\od{{\omega_{\rm d}}}
\def\off{{\omega_{\rm off}}}
\def\fv{{\bf{f}}}
\def\fm{\bf{f}_m}
\def\zh{\hat{z}}
\def\yh{\hat{y}}
\def\xh{\hat{x}}
\def\km{k_{m}}
\def\nh{\hat{n}}

\def\Re{{\rm Re}}
\def\sj{\sum_{j=1}^2}
\def\rk{\rho^{ (k) }}
\def\rek{\rho^{ (1) }}
\def\cek{C^{ (1) }}
\def\rz{\rho^{ (0) }}
\def\rt{\rho^{ (2) }}
\def\rtb{\bar \rho^{ (2) }}
\def\trk{\tilde\rho^{ (k) }}
\def\trek{\tilde\rho^{ (1) }}
\def\trz{\tilde\rho^{ (0) }}
\def\trt{\tilde\rho^{ (2) }}
\def\r{\rho}
\def\tD{\tilde {D}}

\def\s{\sigma}
\def\kb{k_B}
\def\bF{\bar{\cal F}}
\def\F{{\cal F}}
\def\la{\langle}
\def\ra{\rangle}
\def\nn{\nonumber}
\def\up{\uparrow}
\def\dn{\downarrow}
\def\S{\Sigma}
\def\dg{\dagger}
\def\d{\delta}
\def\p{\partial}
\def\l{\lambda}
\def\L{\Lambda}
\def\G{\Gamma}
\def\o{\Omega}
\def\w{\omega}
\def\g{\gamma}
\def\E{{\mathcal E}}

\def\O{\Omega}

\def\vv{ {\bf v}}
\def\jv{ {\bf j}}
\def\jr{ {\bf j}_r}
\def\jd{ {\bf j}_d}
\def\jdd{ { j}_d}
\def\noi{\noindent}
\def\a{\alpha}
\def\d{\delta}
\def\p{\partial} 

\def\la{\langle}
\def\ra{\rangle}
\def\e{\epsilon}
\def\n{\eta}
\def\g{\gamma}
\def\break#1{\pagebreak \vspace*{#1}}
\def\hf{\frac{1}{2}}
\def\rcurs{r_{ij}}

\def\bv{ {\bf b}}
\def\uv{ {\bf u}}
\def\rv{ {\bf r}}
\def\cf{{\mathcal F}}



\title{When does an active bath behave as an equilibrium one? }

\author{Shubhendu Shekhar Khali}
\email{shubhendu@iopb.res.in}
\affiliation{Institute of Physics, Sachivalaya Marg, Bhubaneswar 751005, India}

\author{Fernando Peruani}
\email{fernando.peruani@cyu.fr}
\affiliation{LPTM,
CY Cergy Paris Universit{\'e}, 2 avenue A. Chauvin,
95302 Cergy-Pontoise cedex,
France}
\author{Debasish Chaudhuri}
\email{communicating author: debc@iopb.res.in}
\affiliation{Institute of Physics, Sachivalaya Marg, Bhubaneswar 751005, India}
\affiliation{Max-Planck Institute for the Physics of Complex Systems, N{\"o}thnitzer Str. 38, 01187 Dresden, Germany}

\date{\today}

\begin{abstract}
Active baths are characterized by a non-Gaussian velocity distribution and a quadratic dependence with active velocity $v_0$ of the kinetic temperature and diffusion coefficient.
While these results hold in over-damped active systems, inertial effects lead to normal velocity distributions, with kinetic temperature and diffusion coefficient increasing as 
$\sim v_0^\alpha$ with $1<\alpha<2$. Remarkably, the late-time diffusivity and mobility decrease with mass. Moreover, we show that  the equilibrium Einstein relation is  asymptotically recovered with inertia. In summary, the inertial mass restores an equilibrium-like behavior.  
\end{abstract}

\maketitle



A fluid in equilibrium can be characterized as a heat bath in terms of its temperature, viscous drag, and diffusivity, obeying the Einstein relation~\cite{Chaikin2012}. Active Brownian particles~(ABP), under certain conditions, can constitute a similar homogeneous and isotropic (as opposed to polar) active fluid -- i.e., an active scalar fluid -- in the presence of local energy dissipation and self-propulsion~\cite{Bechinger2016, Marchetti2013, Romanczuk2012, Ramaswamy2019, Gompper2020, Richard2016}. A question that naturally arises is to what extent such a fluid can be characterized as an {\em active} heat bath. The active nature of ABPs is determined by the self-propulsion speed $v_0$ and the orientational diffusivity of the heading direction $D_r$. This leads to a persistent random motion for individual ABPs, in dimension $d$, characterized by a late-time active  diffusivity that scales as $\frac{v_0^2}{D_r d (d-1)}$. Early experiments on tracer particle dynamics in a bacterial suspension  showed enhanced active diffusion~\cite{Wu2000, Gregoire2001}. A higher density reduces diffusivity in equilibrium~\cite{Lahtinen2001} but increases it in a non-equilibrium bacterial bath~\cite{Wu2000}. Various theoretical techniques were used to obtain the impact of an active bath on tracer particles~\cite{Ye2020, Maes2020, Maes2020a, Granek2021a, Rizkallah2022, Demery2014, Demery2011}. Recent works have characterized ABP systems in terms of kinetic temperature, effective diffusivity, and viscous drag as a function of changing activity~\cite{Loi2010, Cugliandolo2019, Mandal2019, Reichert2021, Burkholder2020, Petrelli2020, Caprini2020, Reichhardt2015}. While diffusivity and kinetic temperature increase with activity, a non-monotonic variation of viscous drag has been predicted~\cite{Burkholder2020}. At the motility-induced phase separation (MIPS)~\cite{Fily2012, Redner2013, Caporusso2020}, it was shown that the kinetic temperature could vary across a steady-state system with low (high) temperature characterizing the dense (dilute) phase~\cite{Mandal2019}.   However, it remains unclear to what extent such a description can be developed into a coherent self-consistent picture of active fluid, given, e.g., the breaking of time-reversal symmetry and the absence of equilibrium fluctuation-dissipation.           

Despite tremendous progress in the study of active matter~\cite{Marchetti2013, Bechinger2016, Romanczuk2012, Ramaswamy2019, Gompper2020}, until recently, relatively little attention was paid to the impact of inertia on the active matter except for ~\cite{Scholz2018, Mandal2019, Lowen2020b, Su2021,  Caprini2021, Hecht2022, Solon2022}. 
One reason for this is the extremely short time ($\sim 100$ ns) and length scales (Angstrom) for the ballistic-diffusive crossover in colloidal particles. 
However, for larger active elements, including birds, fish, and animals on the one hand, and artificial macro-sized robots~\cite{Dauchot2019, Scholz2018a, deblais2018boundaries} on the  other hand, inertial effects can be  substantial. This paper considers a homogeneous and isotropic fluid of active Brownian particles (ABP) and probes its active bath-like characteristics.
 A remarkable fact emerges: not only transient behaviors, even the asymptotic properties, including the effective diffusivity and mobility at the steady state, do depend on inertial mass, in sharp contrast to a non-interacting ABP gas. 
Furthermore, the strong non-Gaussian distribution of velocities returns towards equilibrium-like Gaussian for large mass. Finally, with increasing mass, the deviation from the equilibrium fluctuation-dissipation relation drops sharply. 
In summary, while activity amplifies non-equilibrium features, increasing the inertial mass brings the fluid back to equilibrium. 

{\it The model.--} We consider $N$ active Brownian particles (ABP) of mass $m$, a moment of inertia $I$, and diameter $\sigma$ moving in a two-dimensional (2d) rectangular box of area $L_x \times L_y$ with periodic boundary conditions (thus, density $\r=N/L_x L_y$). 
The particles self-propel in  directions $\hat{n}_i=(\cos \theta_i,\sin\theta_i)$ 
with force ${\bf F}_{A,i}=\gamma_t v_0 \hat n_i$. Without interactions, this leads to a propulsion speed $v_0$. 
The heading direction (i.e., $\theta_i$) undergoes a long-time diffusion leading to an effective persistent motion. 
The inertial active dynamics evolve as 
\bea
m\dot{\textbf{v}}_{i} &=& -\gamma_{t}\textbf{v}_{i}+\textbf{F}_{i} +\textbf{F}_{A,i}+ 
\g_t \sqrt{2D_t}\, {\eta_{i}(t)} \nn\\
I\ddot{\theta_{i}} &=& -\gamma_{r}\dot{\theta_{i}} +\g_r \sqrt{2D_r} \,{\zeta_{i}(t)},
\eea
where $\textbf{F}_{i}=- \nabla_i \sum_j U(r_{ij})$, the symbols $\eta_i(t)$ and $\zeta_i(t)$ represent Gaussian white noises, and viscous drags, associated with translation and rotation, are described by $\gamma_{t}\textbf{v}_{i}$ and $\gamma_{r}\dot{\theta_{i}}$, respectively. 
Interactions among particles are due to volume exclusion effects modeled via the Weeks-Chandler-Anderson potential:   
$U(r)=4\epsilon[(\sigma/r)^{12}-(\sigma/r)^{6}] + \epsilon $ if the inter-particle separation $r < r_c=2^{1/6} \, \sigma$ and $U(r)=0$ otherwise. 
The units of length and energy are set by $\sigma$ and $\epsilon$, respectively.
The rotational diffusivity $D_r$ can have a non-thermal active origin. 
Inertial relaxations in translation and rotation take time scales  $\tau_I=m/\gamma_t$ and $\tau_d=I/\gamma_r$. Beyond $\t_d$, diffusion in the heading direction leads to a persistent motion with a bare persistence time $\t_p = 1/D_r$, which sets the unit of time. 
Unless specified otherwise, we use  an equilibrium heat bath with   $D_t=1.0\, \s^2 D_r$, 
and a small and constant 
$\t_d= 0.33 \t_p$~\cite{Mandal2019}.  
We particularly focus on the impact of changing activity in terms of the P{\'e}clet number  ${\rm Pe}=v_0/(D_r\sigma)$ and that of the reduced mass using $M= \tau_I/\tau_p$. This system can  undergo motility-induced phase separation (MIPS) for low enough $M$, and high enough Pe at a moderate density~\cite{Mandal2019}. In order to fully characterize the properties of a homogeneous active bath, we fix the density to a low value, $\r=0.1$, corresponding to a packing fraction $8\%$, such that the system does not phase-separate even in the over-damped limit~\cite{Richard2016}.


\begin{figure}
\centering
\includegraphics[scale=0.68]{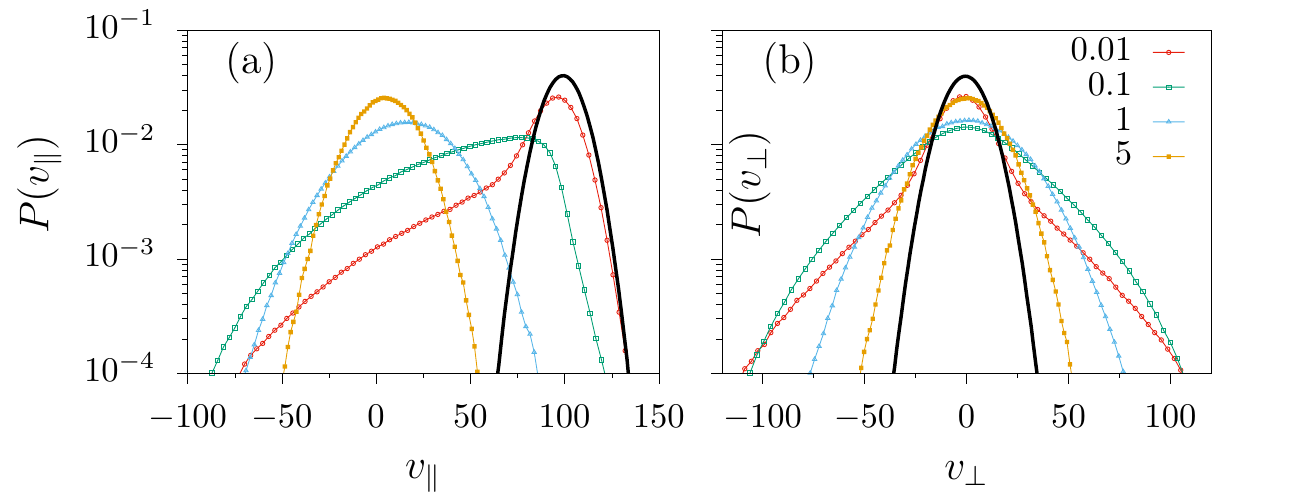}
\caption{{\em Velocity distributions in the particle frame.}  Probability distributions of velocity components (a) parallel $P(v_{\parallel})$  and (b) perpendicular $P(v_{\perp})$ to the heading direction at $Pe=100$ and for different $M$ values indicated in the legend.  Similarly, in $P(v_{\perp})$, non-Gaussian long tails disappear at high inertia. For comparison, we present the distribution functions for non-interacting particles at $M=0.01$ using the solid black lines in both plots.
}
\label{fig:vel_dis_particle_frame}
\end{figure}

\begin{figure}
\includegraphics[scale=0.7]{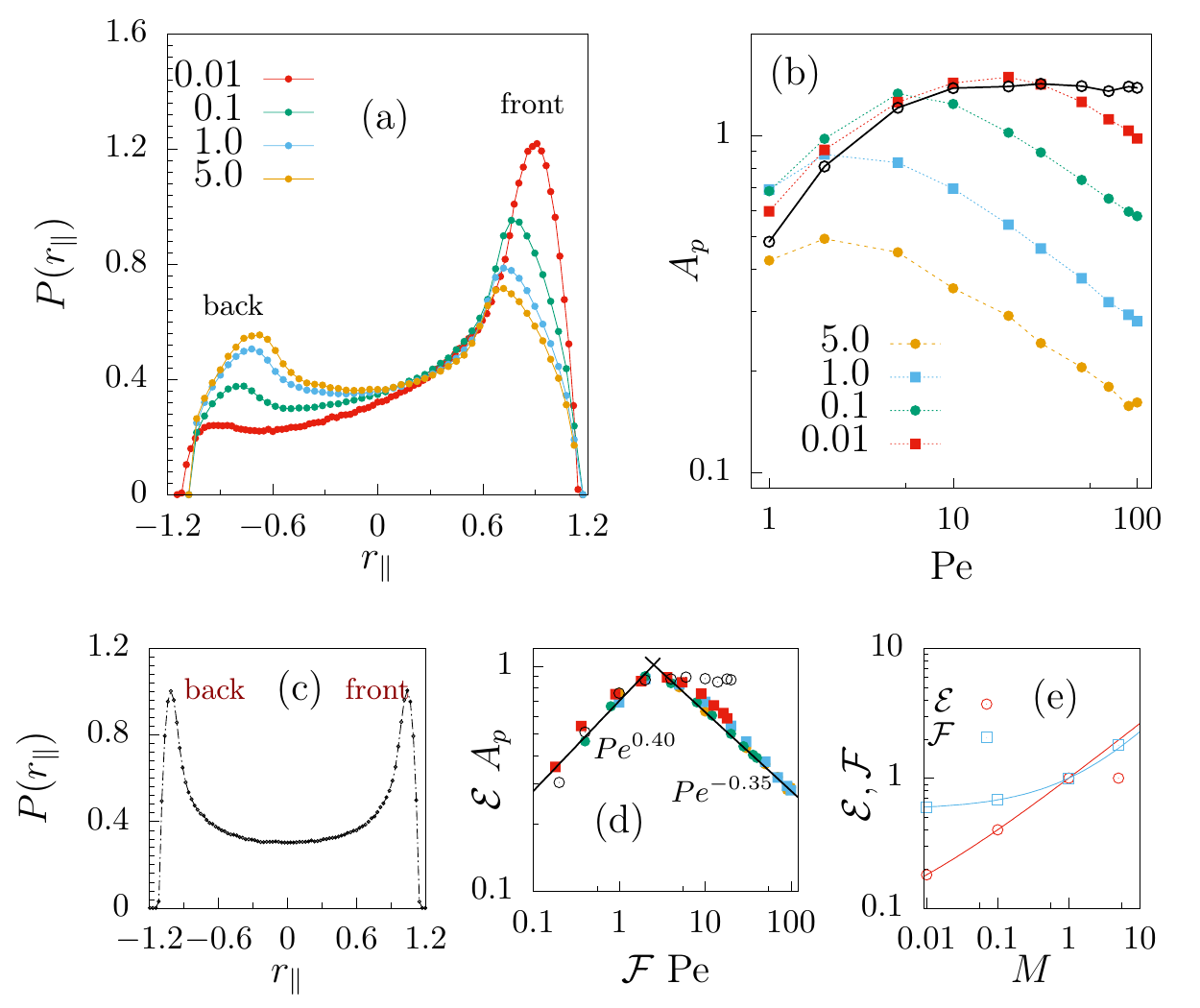}
\caption{{\em Pair distribution in the heading direction:} (a)~$P(r_{\parallel})$ at $Pe=100$ and various $M$ values indicated in the legend. The fore-aft asymmetry decreases with increasing $M$. One gets a fully symmetric distribution at equilibrium, as shown in (c).  
(b)~Variation of asymmetry parameter $A_p$ as a function of Pe is shown for various $M$ values indicated in the legend. 
(d)~An approximate data collapse is obtained for inertial systems using appropriate rescaling of (b).  
(e)~The variation of scale factors shown as a function of $M$. At large $M$, they show approximate dependencies $\mathcal{E} \sim M^{0.4}$ and $ \mathcal{F} \sim M^{0.6}$. 
}
\label{fig:asym_par_variation}
\end{figure}

{\it Results.--} An increase in Pe, as expected, drives the system away from equilibrium. In Fig.\ref{fig:vel_dis_particle_frame}, we show the change in velocity distribution to reveal the impact of inertia. For this purpose,  we consider the two components of the velocity,  in the heading direction $v_\parallel = {\vv \cdot \hat n }$ and perpendicular to it $v_\perp = ({\mathbb 1} - \hat n \hat n) \cdot \vv$. In 2d, $v_\perp$ is a scalar. At small $M$, inertial lag is small. Despite that,  $P(v_\parallel)$ at $M=0.01$ shows a long tail; see the red curve in Fig.\ref{fig:vel_dis_particle_frame}(a). This tail, which is absent in a system of non-interacting active particles, emerges due to enhanced frontal collision in the heading direction and the resultant inertial recoil. At larger inertia, secondary back collisions become prominent, symmetrizing the distribution to a Gaussian-like profile, e.g.,  at $M=5$~\footnote{In the Supplementary Information provided online, we discuss further how $M$ equilibrates the system, analyze the pair distribution function, show the calculation of mobility, diffusivity, and the behavior of scaling exponents describing kinetic temperature, effective diffusivity, and mobility.}. 
The asymmetry in collisions will be scrutinized using a pair distribution function in  Fig.~\ref{fig:asym_par_variation}. The $P(v_\perp)$ distribution also shows inertial restoration of equilibrium-like behavior. The long non-Gaussian tails are observable at small $M$ and disappear with an increase in $M$. The comparison with distributions obtained in the absence of interaction, black curves in Fig.\ref{fig:vel_dis_particle_frame}(a) and (b), shows that the non-Gaussian behaviors are strongly dependent on the inter-particle collisions. At high inertia, bouncing backward and forward from neighboring particles symmetrizes the distributions rendering them equilibrium-like shapes.

\begin{figure*}
\centering
\includegraphics[scale=0.75]{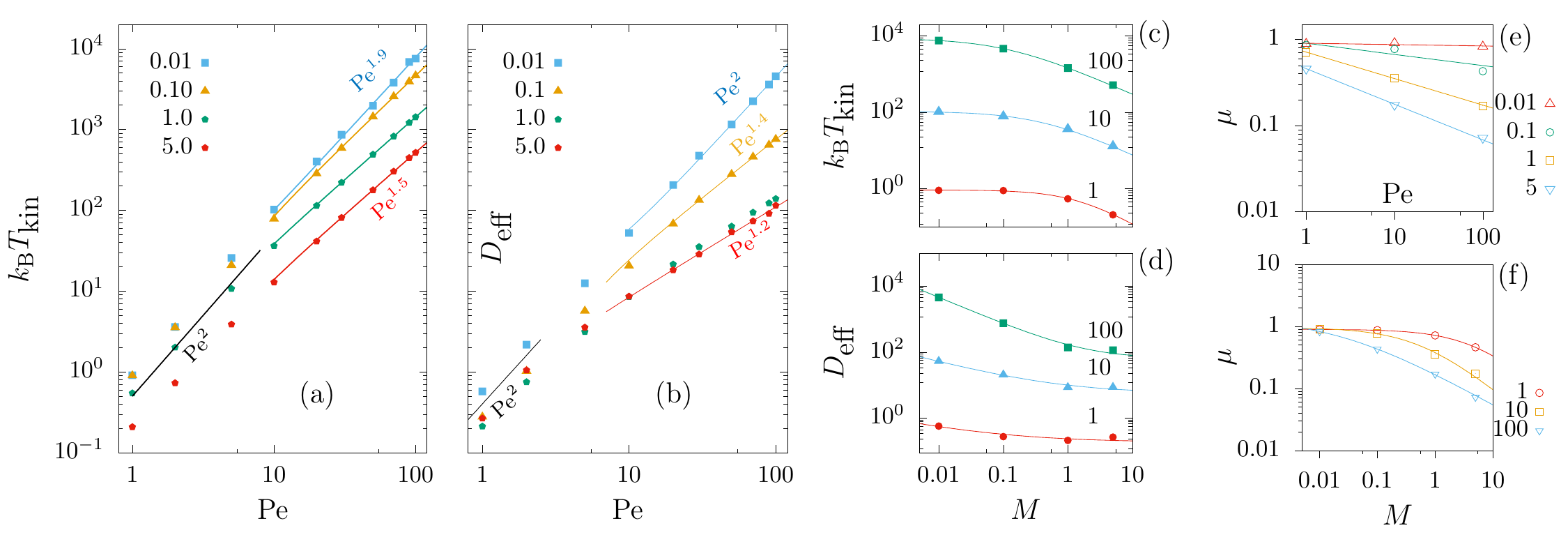}
\caption
{(a)~The increase in kinetic temperature $k_B T_{\rm kin} = m \la v^2 \ra /2$ with Pe is shown at various $M$ values indicated in the legend. At large Pe, $\kb T_{\rm kin} \sim {\rm Pe}^{\a_k}$ with $\a_k=1.90, 1.69, 1.54, 1.53$ at $M=0.01,0.1,1,5$. 
(b)~The increase in effective diffusivity $D_{\textrm {eff}}$ With Pe is shown at at various $M$ values listed in the legend. Asymptotically it grows as $D_{\rm eff} \sim {\rm Pe}^{\a_d}$ with $\a_d=1.98, 1.43, 1.13, 1.17$ at $M=0.01,0.1,1,5$.  
(c)~At large $M$, the kinetic temperature decreases as $\kb T_{\rm kin} \sim M^{-\g_k}$ with $\g_k=1, 0.8, 0.75$  at Pe = 1, 10, and 100. 
(d)~$D_{\rm {eff}}$ reduces with $M$ to saturate, and follows a scaling form $D_{\rm {eff}} = {\cal B} + {\cal D}\, M^{-\g_d}$ with $\g_d=0.43, 0.52, 0.81$ at Pe = 1, 10, and 100. 
(e)~The mobility $\mu$ as a function of Pe decreases as $\mu \sim {\rm Pe}^{-\a_\mu}$ where $\a_\mu=0.02, 0.13,0.31,0.41$ at $M=0.01, 0.1,\, 1, 5$. 
(f)~It decreases with $M$ as $\mu \sim M^{-\g_\mu}$ at large $M$ with $\g_\mu=0.82,0.77,0.52$ at $Pe=1,\, 10,\, 100$.}
\label{fig:kinematics}
\end{figure*}

The impact of interaction manifests in the pair-distribution function in the heading direction, i.e., the probability $P(r_\parallel)$ of finding a neighboring particle in direction $\hat{n}$; see Fig.\ref{fig:asym_par_variation}. Due to enhanced frontal collision, more particles accumulate in front.  Such accumulations in front and associated depletion wakes have been analyzed recently in overdamped  dilute ABPs~\cite{Poncet2021}. A peak in the back appears due to secondary back collisions experienced by inertial ABPs after frontal recoil. With increasing $M$, recoil increases, reducing the frontal accumulation and increasing the secondary collisions from the back -- these affect the symmetrization of the pair distribution -- restoring equilibrium-like behavior. 

The asymmetry in the pair distribution is quantified in terms of the  parameter $A_p$ that measures the difference between the heights of the front and back peaks~(Fig.\ref{fig:asym_par_variation}(b)\,). For fixed $M$, the asymmetry initially increases with Pe.  
For overdamped systems (black solid line denoting $M=0$), the increase is followed by saturation in the absence of recoil.  In contrast, in the presence of inertia, after the initial increase,  $A_p$ decreases with Pe.
The data collapse in Fig.\ref{fig:asym_par_variation}(d) shows that at small Pe, the asymmetry increases with Pe as $A_p\sim {\rm Pe}^{0.4}$ and then decreases as $A_p \sim {\rm Pe}^{-0.35}$. Such scaling properties are common to all inertial ABPs. The increase is due to enhanced frontal collisions associated with increased activity. The decrease is due to inertial recoil and is the reason behind the restoration of equilibrium-like behavior.

Another way of measuring how far the system is from equilibrium is given by  the extent to which the equilibrium Einstein relation is violated ${\cal I} = \mid D_{\rm eff}- \mu\, k_B T_{\rm kin} \mid$, where $\mu$ is the particle mobility and $T_{\rm kin}$ the so-called kinetic energy~\cite{Blickle2007, Chaudhuri2012}.  
Following a tracer particle dynamics is a useful tool to characterize the properties of a bath~\cite{Ye2020, Maes2020, Granek2021a, Rizkallah2022, Demery2014, Demery2011}.
In the presence of translational fluctuations, the impact of the activity on diffusivity at low Pe gets overshadowed by $D_t$. While these effects can be subtracted out, for simplicity and intending to understand the impact of the activity, in the following, we set $D_t=0$. 
From the late-time behavior of the mean-squared displacements (MSD), 
we obtain the effective diffusivity $D_{\rm eff}$. The kinetic temperature $k_B T_{\rm kin}$ is readily obtainable from the velocity fluctuations. Using a separate numerical calculation of the change in velocity $\la v_x\ra$ of a test particle under an external force $f_x$, we obtain the mobility $\mu = (\p \la v_x \ra/\p f_x)_{f_x=0}$ around the non-equilibrium steady states of the ABPs. 
In the presence of an external force on the test particle, a local statistical reorganization of other ABPs follows. Such a reorganization depends on active speed and inertia, resulting in mobility variation. Using these, we obtain violation ${\cal I}$.

In Fig.\ref{fig:kinematics}, we show the variations of kinetic temperature, diffusivity, and mobility as a function of Pe and $M$. At large enough values of Pe and $M$, and they show the following scaling forms: 
{
(i)~$\kb T_{\rm kin} \sim {\rm Pe}^{\a_k}$ and $\kb T_{\rm kin} \sim M^{-\g_k}$,  
(ii)~$D_{\rm eff} \sim {\rm Pe}^{\a_d}$ and $D_{\rm eff}  \sim M^{-\g_d}$,
(iii)~$\mu \sim {\rm Pe}^{-\a_\mu}$ and $\mu \sim M^{-\g_\mu}$. Note that mobility decreases both with Pe and $M$, while the other quantities increase with Pe. 
}

The kinetic temperature $k_B T_{\rm kin} = m \la v^2 \ra /2$ increases as a function of Pe~(Fig.\ref{fig:kinematics}(a)\,). The black solid line captures a Pe$^2$ scaling common to all $M$ values at small Pe. The asymptotic behavior $\kb T_{\rm kin}\sim {\rm Pe}^{\a_k}$  at various $M$ values shows how the exponent $\a_k$ decreases with $M$. 

In Fig.\ref{fig:kinematics}(b) we show how the effective diffusivity $D_{\textrm {eff}} \sim {\rm Pe}^{\a_d}$ increases with Pe. At small Pe, $D_{\textrm {eff}} \sim {\rm Pe}^2$ for all $M$. This behavior continues to all Pe in the limit of small inertia, $M=0.01$, as in over-damped systems. However, with increasing inertia, the value of the growth exponent $\a_d$ decreases, and finally, $D_{\rm {eff}} \sim {\rm Pe}^{\a_d}$ becomes approximately linear with Pe at $M \gtrsim 1$.

In Fig.\ref{fig:kinematics}(c) and (d), we show the $M$-dependency of the kinetic temperature and effective diffusivity. The kinetic temperature fits to the functional form $\kb T_{\rm kin} = {\cal A}/({\cal C} + M^{\g_k})$, such that  at large $M$, they show a scaling form $\kb T_{\rm kin} \sim M^{-\g_k}$ with the decay exponent $\g_k$ decreasing with Pe. 
The effective diffusivity $D_{\rm {eff}}$ decays with $M$ until it saturates, and follows a scaling form $D_{\rm {eff}} = {\cal B} + {\cal D}\, M^{-\g_d}$ with the decay exponent $\g_d$ increasing with Pe. 

Finally, we consider the mobility $\mu$. In the over-damped limit, $\mu$ is approximately independent of activity Pe~($M=0.01$ in Fig.\ref{fig:kinematics}(f)\,). This behavior agrees with the observation in Ref.~\cite{Petrelli2020} at low densities. However, in the presence of inertia, mobility decreases with both Pe and $M$. At  the larger activity, an  increase in frontal collision frequency in the direction of external force increases the effective drag, leading to an effective thickening of the active fluid. 
Such a thickening is in qualitative agreement with Ref.~\cite{Reichhardt2015}, unlike the non-monotonic variation predicted in Ref.~\cite{Burkholder2020}, and oppose to the active thinning predicted in Ref.~\cite{Jayaram2023}.    
As a function of Pe, the mobility  decreases as $\mu \sim {\rm Pe}^{-\a_\mu}$  where the exponent $\a_\mu$ itself increases with  $M$~(Fig.\ref{fig:kinematics}(f)\,). 
The collision and recoil in inertial ABPs further increase the drag coefficient for  higher inertial mass. The reduction of $\mu$ with $M$ follows a  form similar to that of the dependence of $\kb T_{\rm kin}$ on $M$~(Fig.\ref{fig:kinematics}(c)\,), such that at large $M$  $\mu \sim M^{-\g_\mu}$. Note that $\g_\mu$ also decreases with  $Pe$.  

\begin{figure}
\centering
\includegraphics[scale=0.7]{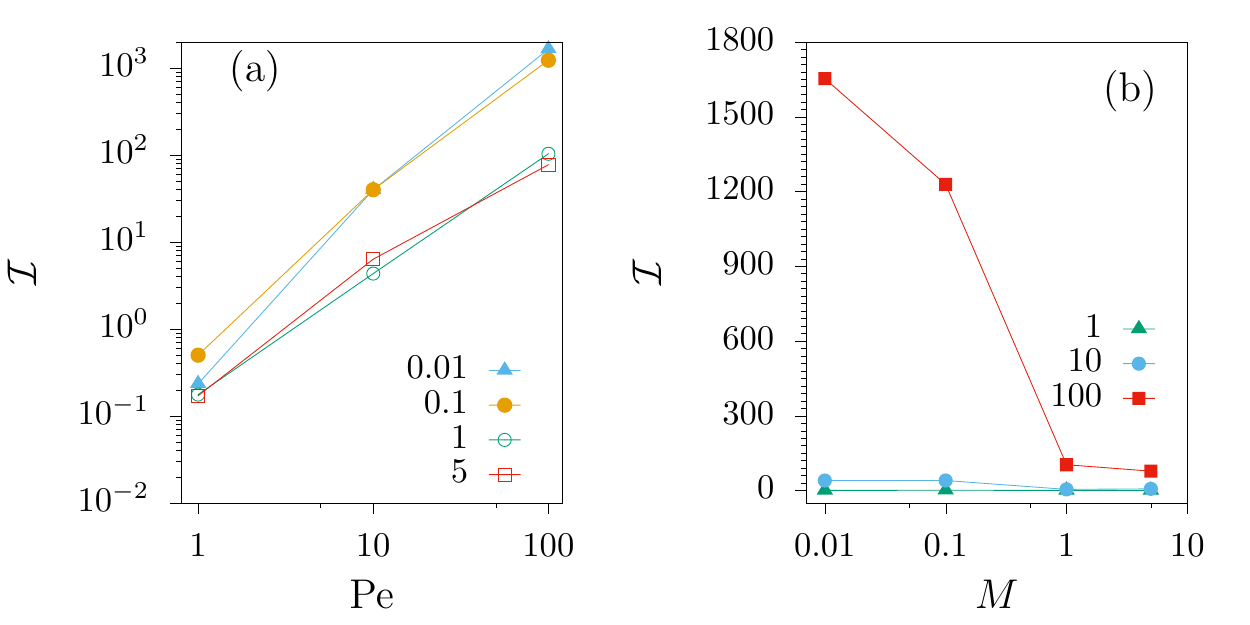}
\caption{ The violation of equilibrium fluctuation-dissipation  ${\cal I}= \mid D_{\textrm {eff}}-\mu \kb T_{\rm {kin}} \mid$, a measure of how far  the system is from equilibrium as a function of $Pe$ and $M$, is shown in (a) and (b), respectively.  ${\cal I}$ increase with Pe as Pe$^{\a_d}$ or Pe$^{(\a_k-\a_\mu)}$. With $M$, it decreases to saturate at ${\cal B}$.}
\label{fig:violation}
\end{figure}

With the help of these results, we find that the violation ${\cal I}$ increases with Pe, but decreases with $M$. In particular, our findings indicate that ${\cal I} \sim \mid {\rm Pe}^{\a_d}-{\rm Pe}^{-\a_\mu +\a_k}\mid$, i.e., it can grow as ${\rm Pe}^{\a_d}$ or ${\rm Pe}^{\a_k-\a_\mu}$ depending on whether $\a_d>(\a_k-\a_\mu)$ or not~(Fig.\ref{fig:violation}(a)\,). On the other hand, ${\cal I} \sim \mid {\cal B} + {\cal D} M^{-\g_d}-M^{-(\g_\mu +\g_k)}\mid$ decreases with $M$ to saturate at a Pe-dependent value ${\cal B}$~(Fig.\ref{fig:violation}(b)\,). Such a decrease is more significant at higher Pe. For small Pe, the system remains close to equilibrium for all $M$ values.  

{\it Concluding remarks.-- }
We have shown how the non-equilibrium properties of an active fluid consisting of ABPs, at a density far away from the onset of MIPS behave as a function of activity and inertial mass. While the departure from equilibrium gets pronounced with increasing Pe, i.e., activity, we found that inertial mass restores equilibrium-like properties. 
In particular, we showed that the non-Gaussian velocity distributions,  the fore-aft asymmetry in the (heading-direction) pair distribution, and the absence of an Einstein fluctuation-dissipation relation between diffusivity, mobility, and kinetic temperature observed in active over-damped systems, crossover to their equilibrium counterparts with inertial mass. In short, we found that the inertial recoil can effectively thermalize the active fluid. 
Remarkably, the late-time diffusivity and mobility of the bath depend on inertial mass, in contrast to free ABPs. 
The effective diffusivity and temperature grow with active velocity and decrease with inertial mass. In contrast, effective mobility decreases with mass and activity. Together, these findings show a reduction in the violation of equilibrium fluctuation-dissipation with increasing inertia. In summary, inertia brings back equilibrium-like behavior in the active fluid.     

\begin{acknowledgements}
Numerical calculations were supported by SAMKHYA, the high-performance computing facility at the Institute of Physics, Bhubaneswar. D.C. thanks Arghya Majee for critical reading of the manuscript, C.Y. Cergy Paris Universit{\'e} for a Visiting Professorship which enabled the initiation of current work, SERB, India, for financial support through Grant No. MTR/2019/000750,  and International Centre for Theoretical Sciences (ICTS-TIFR), Bangalore, for an Associateship. 
F.P. acknowledges financial support from C.Y. Initiative of Excellence (grant 'Investissements d'Avenir' ANR-16-IDEX-0008), INEX 2021 Ambition Project CollInt and Labex MME-DII, projects 2021-258 and 2021-297. 
\end{acknowledgements}


%


\onecolumngrid
\appendix 

\section{Supplemental Information: When does an active bath behave as an equilibrium one? }

\begin{figure}[h!]
\centering
\includegraphics[scale=0.75]{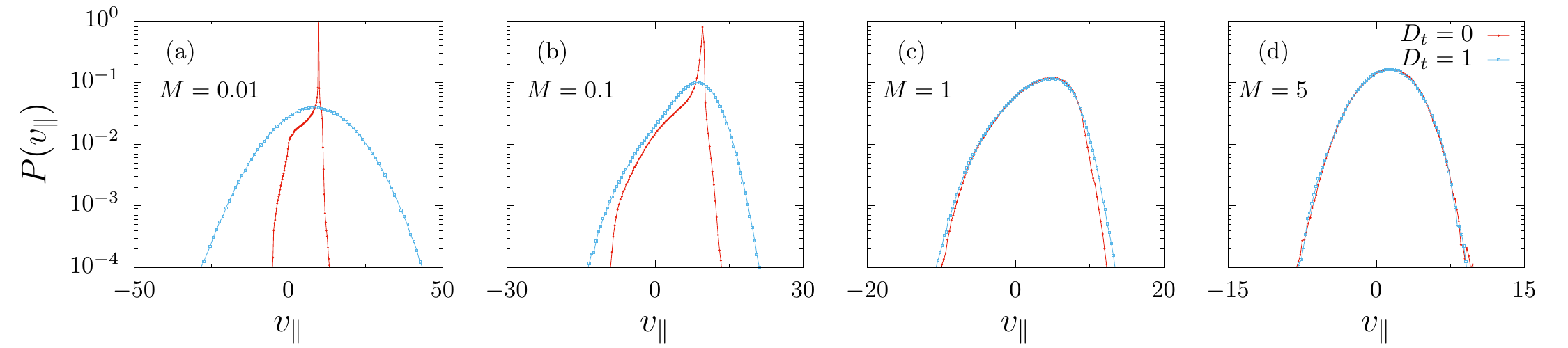}
\caption{Inertia brings back equilibrium-like features. The speed distributions in heading direction at $Pe=10$ for different masses $M = 0.01$ in (a), $M = 0.1$ in (b), $M = 1$ in (c) and $M = 5$ in (d). In each plot, blue and red curves represent the results for $D_t$ = 1.0 and 0.0, respectively.}
\label{fig:pv_tb}
\end{figure}

\subsection{Inertial thermalization}
\label{sec_thermal}
As has been pointed out in the main text, with increasing $M$, the velocity distribution gets thermalized due to collisions and recoil. Here we supplement that finding with an independent measure in which we compare the velocity distribution in the heading direction in the presence ($D_t=1$) and absence ($D_t=0$) of a thermal bath~(Fig.\ref{fig:pv_tb}). 
For small $M$, the two distributions are significantly different from each other. With increasing $M$, the distribution in the absence of thermal bath starts to {\em thermalize}, aided by relatively faster orientational relaxation of the heading direction during a slow inertial relaxation, and collisions from the front and secondary collisions from back after the frontal recoil. As a result, they start to come close to each other to merge at $M\gtrsim 1$.  

\begin{figure}[h!]
\includegraphics[scale=0.75]{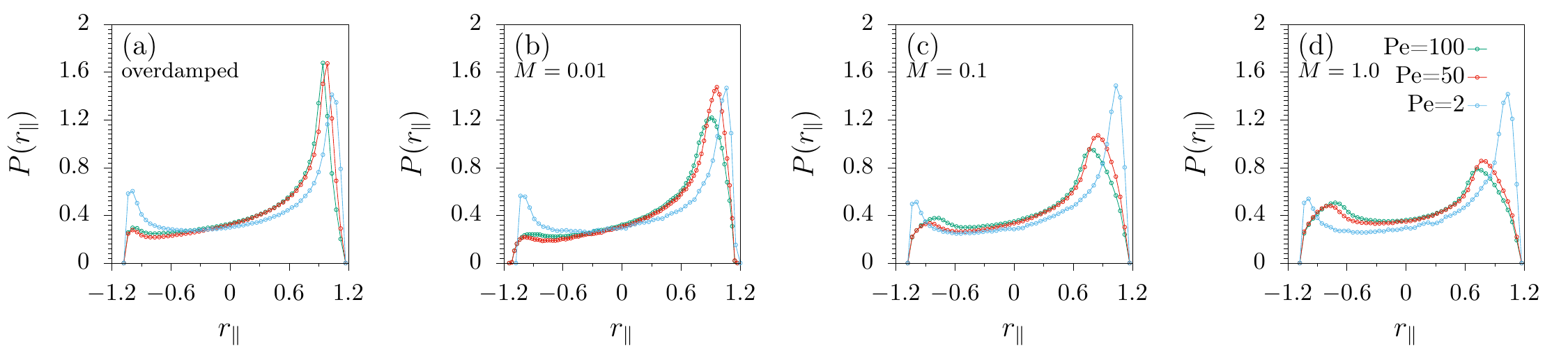}
\caption{ Probability distribution of particle separation of interacting neighbors projected along the heading direction $P(r_{\parallel})$ 
at the system density $\rho=0.1$ for $Pe= 2, 50$ and $100$. The results In (a) correspond to an over-damped system and in (b), (c), and (d) for an under-damped system with $M$ values = 0.01, 0.1, and 1.0, respectively. The common legend representing different $Pe$ values is placed inside plot (d).
\label{fig:par_sep_dis}
}
\end{figure}

\subsection{Pair distribution}
In Fig.\ref{fig:par_sep_dis}, we show the distribution of particles around a test particle in its heading direction at Pe=1, 10, and 100 and inertia $M=0,0.01,0.1, 1$. These distributions are used to calculate the asymmetry parameter $A_p$ presented in the main text. 

\begin{figure}[h]
\includegraphics[scale=0.75]{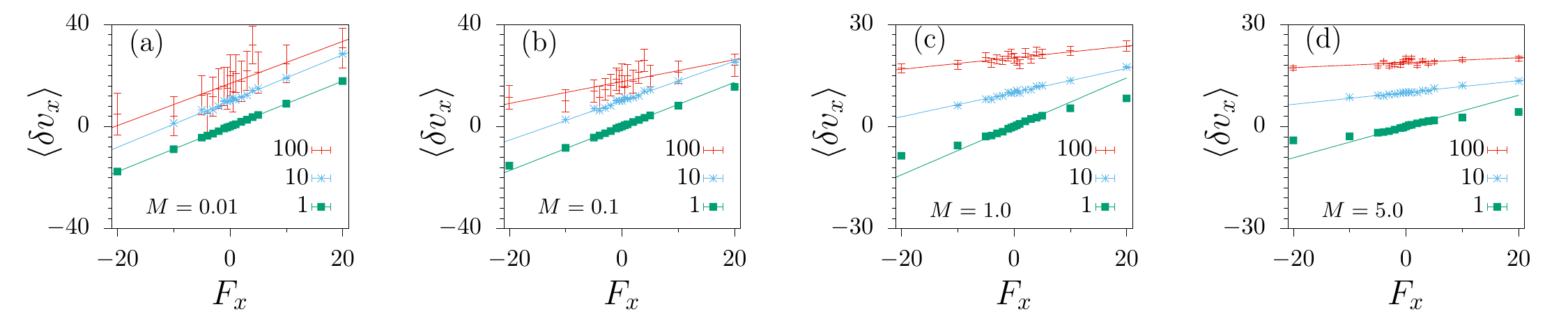}
\caption{ 
The mean velocity of a tagged particle in response to the external force $F_x$ is shown for $Pe=1$, 10, and 100 in (a) at $M = 0.01$, in (b) at $M = 0.1$, in (c) at $M = 1.0$ and in (d) at $M = 5$. The slope of the corresponding linear fits at $F_x=0$ gives mobility $\mu= \lim_{F_x \to 0} \la \delta v_x \ra /F_x$. By definition $\la \d v_x\ra=0$ at $F_x=0$.  For the sake of clear visualization, each curve for Pe = 10 and 100 is shifted upwards.
\label{fig:vel_response}
}
\end{figure}

\subsection{\label{sec:mobility_extraction} Mobility}
To measure mobility, we apply an additional force $F_x$ on the tagged particle and calculate its mean velocity along the same direction $\langle v_x \rangle$ for various $F_x$ values. The mean velocity $\langle \delta v_x \rangle = \langle v_x \rangle|_{F_x}  -\langle v_x\rangle|_{F_x=0}$ as a function of $F_x$ is shown in the figure \ref{fig:vel_response}. The slope of the linear fit for $\langle \delta v_x \rangle$ vs $F_x$ curve near $F_x=0$ gives mobility. In the overdamped regime $M=0.01$, mobility does not show dependency on Pe as its value is observed to be the same $\mu \sim 0.9$ for Pe = 1, 10, and 100 Fig.\ref{fig:vel_response} (a). In the underdamped limit ($M \geq 0.1$), mobility decreases with Pe, see Fig.\ref{fig:vel_response} (b),(c), (d).

\begin{figure}[h!]
\includegraphics[scale=0.75]{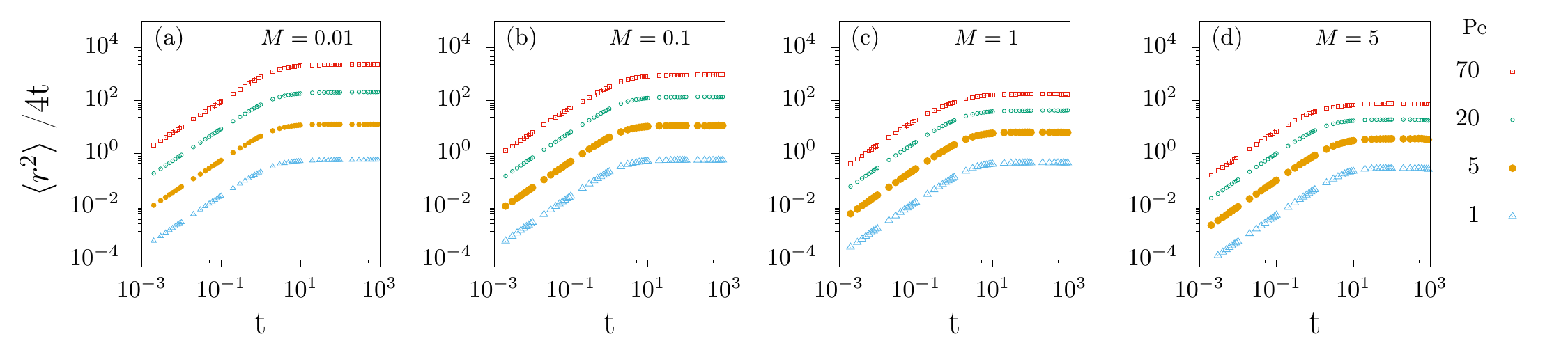}
\caption{
Mean square displacement for Pe = 1, 5, 20, and 70 in (a) at $M = 0.01$, in (b) at $M = 0.1$, in (c) at $M = 1.0$ and in (d) at $M = 5$. The common legend representing different Pe values is placed at the extreme right. These calculations are done for $D_t=0$.
\label{fig:msd}
}
\label{fig:msd}
\end{figure}

\subsection{Mean-squared displacement}
\label{sec:msd}
In Fig.\ref{fig:msd}, we plot mean-squared displacement $\la r^2\ra$ scaled by time $t$ for different $M$ and Pe values. All of them show ballistic to diffusive crossover at a time scale determined by the orientational persistence of the heading direction, a quantity that is kept constant in this paper. The asymptotic diffusivities $D_{\rm eff} = \lim_{t\to \infty} \la \rv^2\ra/4t$ are obtained from these graphs and used in the main text. 

\begin{figure}[h!]
\centering
\includegraphics[scale=0.7]{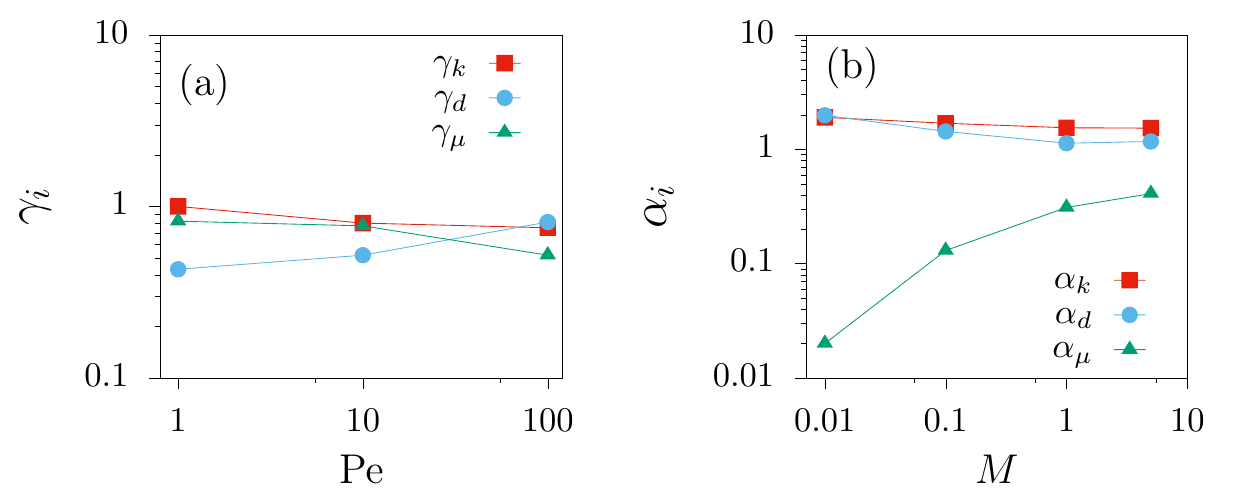}
\caption{ Scaling exponents $\gamma_i$ and  $\alpha_i$ used in Fig.4 are presented as a function of Pe~(a) and $M$~(b), respectively. $i=k, d, \mu$ represent exponents of kinetic energy, diffusion, and mobility.}
\label{fig:exponents}
\end{figure}

\subsection{Scaling exponents}
\label{sec_exponent}
The variations of scaling exponents determining the Pe and $M$ dependence of kinematic temperature, effective diffusivity, and mobility are shown in Fig.\ref{fig:exponents}.

\nocite{*}

\end{document}